\documentclass[%
reprint,
superscriptaddress,
nofootinbib,
 amsmath,amssymb,
 aps,
]{revtex4-1}

\usepackage{graphicx}
\usepackage{dcolumn}
\usepackage{bm}

\usepackage{xcolor}
\usepackage[normalem]{ulem}

\usepackage[pagebackref=false]{hyperref}

\begin{document}

\title{Dispersive sensing of charge states in a bilayer graphene quantum dot}

\author{L. Banszerus}
\email{luca.banszerus@rwth-aachen.de.}
\affiliation{JARA-FIT and 2nd Institute of Physics, RWTH Aachen University, 52074 Aachen, Germany,~EU}%
\affiliation{Peter Gr\"unberg Institute  (PGI-9), Forschungszentrum J\"ulich, 52425 J\"ulich,~Germany,~EU}

\author{S. M\"oller}
\author{E.~Icking}
\affiliation{JARA-FIT and 2nd Institute of Physics, RWTH Aachen University, 52074 Aachen, Germany,~EU}%
\affiliation{Peter Gr\"unberg Institute  (PGI-9), Forschungszentrum J\"ulich, 52425 J\"ulich,~Germany,~EU}

\author{C.~Steiner}
\affiliation{JARA-FIT and 2nd Institute of Physics, RWTH Aachen University, 52074 Aachen, Germany,~EU}%

\author{D.~Neumaier}
\affiliation{AMO GmbH, Gesellschaft f\"ur Angewandte Mikro- und Optoelektronik, 52074 Aachen, Germany, EU}
\affiliation{University of Wuppertal, 42285 Wuppertal, Germany, EU}

\author{M.~Otto}
\affiliation{AMO GmbH, Gesellschaft f\"ur Angewandte Mikro- und Optoelektronik, 52074 Aachen, Germany, EU}

\author{K.~Watanabe}
\affiliation{Research Center for Functional Materials, 
National Institute for Materials Science, 1-1 Namiki, Tsukuba 305-0044, Japan
}
\author{T.~Taniguchi}
\affiliation{ 
International Center for Materials Nanoarchitectonics, 
National Institute for Materials Science,  1-1 Namiki, Tsukuba 305-0044, Japan
}%

\author{C. Volk}
\author{C. Stampfer}
\affiliation{JARA-FIT and 2nd Institute of Physics, RWTH Aachen University, 52074 Aachen, Germany,~EU}%
\affiliation{Peter Gr\"unberg Institute  (PGI-9), Forschungszentrum J\"ulich, 52425 J\"ulich,~Germany,~EU}%

\date{\today}


\begin{abstract} 
We demonstrate dispersive readout of individual charge states in a gate-defined few-electron quantum dot in bilayer graphene. 
We employ a radio frequency reflectometry circuit, where an LC resonator with a resonance frequency close to 280~MHz is directly coupled to an ohmic contact of the quantum dot device.
The detection scheme based on changes in the quantum capacitance operates over a wide gate-voltage range and allows to probe excited states down to the single-electron regime.
Crucially, the presented sensing technique avoids the use of an additional, capacitively coupled quantum device such as a quantum point contact or single electron transistor, making dispersive sensing particularly interesting for gate-defined graphene quantum dots. 
\end{abstract}

\maketitle

The detection of charge transitions in quantum dots (QDs) is a key ingredient in solid state quantum computation. A wide-spread method is the use of capacitively coupled quantum point contacts (QPCs) or single-electron transistors (SETs) 
acting as charge sensors. In semiconductor devices, commonly the conductance of the sensor is either measured directly by transport~\cite{Elzerman2003Apr,Ihn2009Sep} or indirectly via the reflectance of an impedance matched LC resonant circuit connected to the charge sensor~\cite{Schoelkopf1998May,Reilly2007Oct}. 
This technique has enabled single-shot detection of charge and spin states~\cite{Reilly2007Oct,Elzerman2004Jul,Barthel2010Apr,Prance2012Jan,Volk2019Aug} and has been used in state-of-the-art spin qubit devices~\cite{Yoneda2017Dec,Watson2018Feb,Zajac2018Jan,Huang2019Apr}.

An alternative charge detection scheme is dispersive readout, a technique based on a resonant circuit, i.e. tank circuit, which is directly connected either to one of the reservoirs of the QD device or to one of the gates controlling the dot. This method detects the change in quantum capacitance occurring when a QD is in resonance with the reservoirs and hence an electron can tunnel in or out. Dispersive readout schemes avoid the need of a close-by charge sensor and have been applied to QD systems in GaAs~\cite{Petersson2010Aug,Colless2013Jan}, silicon~\cite{Gonzalez-Zalba2015Jan} and carbon nanotubes~\cite{Chorley2012Jan}. Over the past years, the bandwidth of dispersive readout schemes has been improved and single-shot spin readout has been demonstrated recently~\cite{Urdampilleta2019Aug,West2019May,Zheng2019Aug}.

Bilayer graphene (BLG) is an attractive host material for spin qubits due to its small spin-orbit and hyperfine interaction, promising long spin coherence times~\cite{Trauzettel2007Feb}.
The development of clean van der Waals heterostructures, where a BLG sheet is encapsulated into hexagonal boron nitride (hBN)~\cite{Wang2013Nov} and a graphite crystal is used as a back gate~\cite{Overweg2018Jan,Banszerus2020May}, has lead to a boost in device quality and has enabled the implementation of QDs~\cite{Eich2018Jul,Kurzmann2019Jul,Banszerus2020Oct} and double QDs~\cite{Banszerus2018Aug,Banszerus2020Mar}.
While a DC charge detector in gate-defined BLG QDs has been recently demonstrated~\cite{Kurzmann2019Aug}, these experiments required the measurement of a current through a capacitively coupled single electron transistor or QPC. In contrast, radio frequency (RF) reflectometry based detection schemes have not yet been realized in graphene and BLG. 

In this Letter, we demonstrate dispersive charge sensing on a few-electron bilayer graphene quantum dot. We measure the reflectance of an LC resonator directly connected to an ohmic contact of the gate-defined BLG QD device. Our detection scheme based on the change of the quantum capacitance allows to resolve the excited state spectrum of a QD with an electron occupation between zero and ten. Furthermore, we characterize the signal-to-noise ratio as a function of RF frequency, RF power and the measurement bandwidth. Dispersive readout techniques are particularly interesting for BLG QD devices, as the small, voltage controlled band gap renders the implementation of an additional mesoscopic electrometer particularly challenging in this material.

\begin{figure}[]
	\centering
\includegraphics[draft=false,keepaspectratio=true,clip,width=1\linewidth]{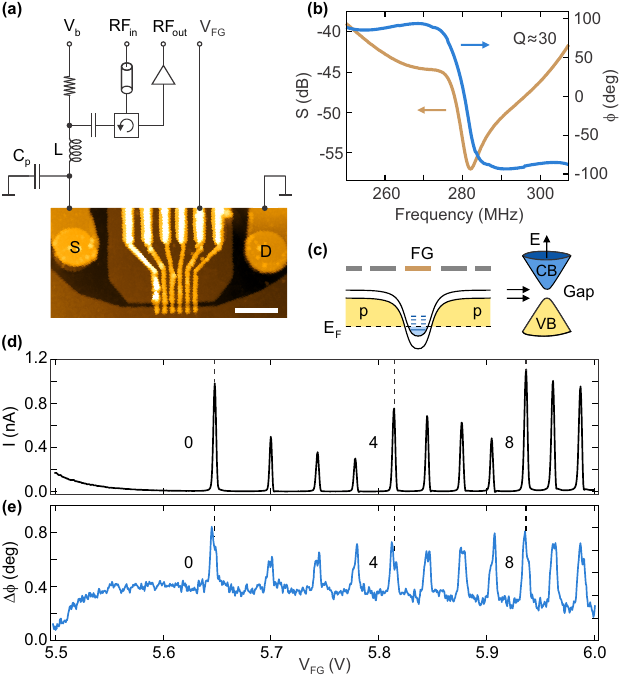}
\caption[Fig01]{  
\textbf{(a)} Atomic force micrograph of a representative device. The scale bar measures 500~nm. The ohmic contacts to the BLG sheet are labelled source (S) and drain (D). The split gates (SGs) define the conducting channel, which can be modulated by voltages applied to the finger gates (FG). A tank circuit formed by the inductor $L$ (820~nH) on the PCB and the parasitic capacitances $C_\mathrm{p}$ of the PCB and the bond wire is connected to the source contact while the drain is kept at ground potential. The RF carrier is coupled in via attenuators and a circulator. The DC bias voltage $V_\mathrm{b}$ is applied via a bias tee. The RF signal reflected by the resonant circuit is amplified at 4~K.
\textbf{(b)} Magnitude $S$ (orange trace) and phase $\phi$ (blue trace) of the RF signal reflected off the tank circuit as a function of the frequency $f$.
\textbf{(c)} Schematic of the band edge profile along the channel between the source and the drain region.
\textbf{(d)} Conductance through a QD formed under the finger gate in the few electron regime as a function of the gate voltage $V_\mathrm{FG}$ at $V_\mathrm{b} = 200~\mu$V. Numbers indicate the electron occupation in the regions of Coulomb blockade.
\textbf{(e)} Simultaneously measured phase shift $\Delta\phi$ of the reflected RF signal as a function of $V_\mathrm{FG}$ ($f = 280$~MHz).
}
\label{f1}
\end{figure}

Fig.~\ref{f1}(a) shows an atomic force micrograph of a fabricated device together with the RF reflectometry setup. 
The device consists of a BLG flake, which has been encapsulated between two crystals of hBN of approximately 25~nm thickness using conventional dry van-der-Waals stacking techniques~\cite{Engels2014Sep,Wang2013Nov}. The heterostructure is placed on a graphite flake, acting as a back gate~\cite{Banszerus2018Aug}. On top of this stack, Cr/Au split gates with a lateral separation of 130~nm are deposited. Separated from the split gates (SGs) by a $25 \, \mathrm{nm}$ thick layer of atomic layer deposited $\mathrm{Al_2O_3}$, we fabricated $90 \, \mathrm{nm}$ wide finger gates (FGs) with a pitch of $150 \, \mathrm{nm}$.

In order to perform experiments in the radio frequency domain, the sample is mounted on a printed circuit board (PCB) which is equipped with low pass filtered DC lines (1~nF capacitors to ground) as well as unfiltered 50~$\Omega$ impedance matched AC lines. 
The source contact of the device is bonded to a tank circuit formed by a surface-mount inductor ($L=820$~nH) on the PCB and the parasitic capacitances $C_\mathrm{p}$ of the PCB. 
The RF signal is coupled to the tank circuit via attenuators (-36~dB) inside the cryostate, a circulator (Quinstar QCY-002003VM00) and a bias-tee allowing also for the application of a DC bias voltage $V_\mathrm{b}$. The signal reflected off the tank circuit is amplified by a cryogenic amplifier (Quinstar QCA-U-230-30HZ1) and demodulated to DC at room temperature using a Zürich instruments UHFLI lock-in amplifier.
The drain contact is bonded to an RF ground. All gates are bonded to low-pass filtered DC lines to reduce charge noise.
All measurements are performed in a $^3$He/$^4$He dilution refrigerator at a base temperature of around 10~mK and an electron temperature of around 60~mK. 

\begin{figure}[]
	\centering
\includegraphics[draft=false,keepaspectratio=true,clip,width=\linewidth]{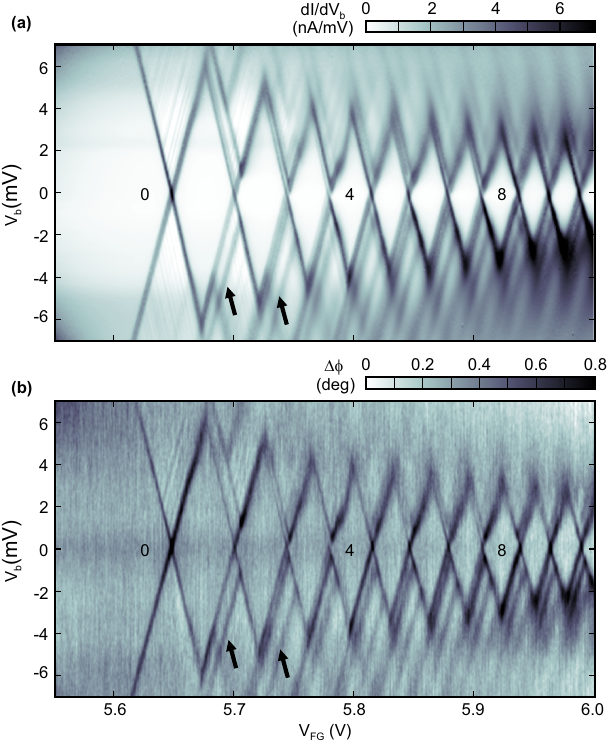}
\caption[Fig02]{  
\textbf{(a)} Finite bias spectroscopy measurements showing the differential conductance through the QD in the few electron regime. The derivative with respect to the bias axis has been performed numerically.
\textbf{(b)} Simultaneously recorded dispersive measurement showing the phase shift $\Delta\phi$ of the reflected RF carrier. A spectrum of excited state transitions can be identified in both measurements, two prominent states are highlighted by the arrows.
}
\label{f2}
\end{figure}
\begin{figure*}[]
	\centering
\includegraphics[draft=false,keepaspectratio=true,clip,width=\linewidth]{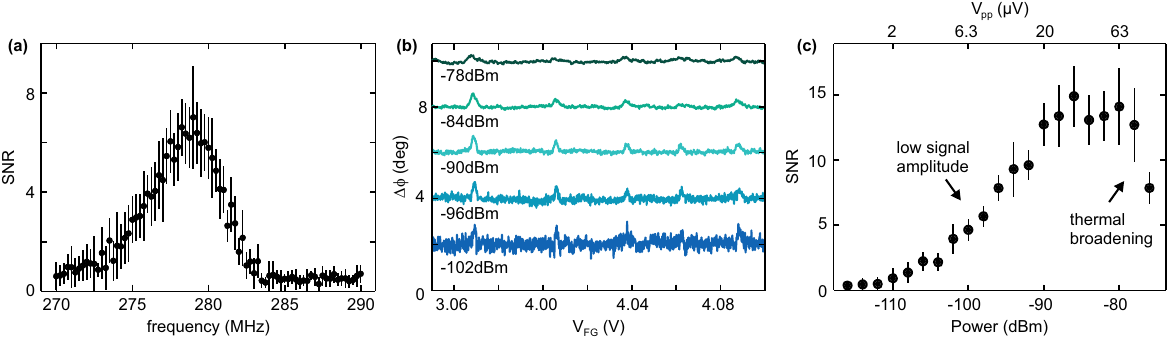}
\caption[Fig03]{
\textbf{(a)}  Signal-to-noise ratio (SNR) of the phase shift $\Delta\phi$ of the single electron Coulomb peak as a function of $f$. A RF power of $-95$~dBm was applied to the resonant circuit. The measurement bandwidth was set to $20$~Hz.
\textbf{(b)}  Phase shift of the reflected signal as a function of $V_\mathrm{FG}$ for different RF power at the sample ($-36$~dB cryogenic attenuators taken into account; $f = 279$~MHz). Curves are offset for clarity.
\textbf{(c)} SNR of the phase response as a function of the RF power at $f = 279$~MHz.
}
\label{f3}
\end{figure*}
\begin{figure}[]
	\centering
	\includegraphics[draft=false,keepaspectratio=true,clip,width=0.8\linewidth]{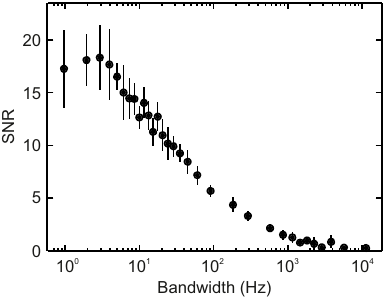}
	\caption[Fig04]{
		Signal-to-noise ratio (SNR) of the phase response as a function of the measurement bandwidth at $f$ = 280~MHz and -90~dBm.
	}
	\label{f4}
\end{figure}

Fig.~\ref{f1}(b) shows the magnitude $S$ and the phase $\phi$ of the reflected RF response as a function of the frequency~$f$. A resonance frequency of $f_0 = 279$~MHz can be determined, which corresponds to a total capacitance of $C_\mathrm{tot}\approx 0.4~$pF in a simplified LC tank circuit with the resonance frequency $f_0 = 1/(2\pi\sqrt{LC_\mathrm{tot}})$ neglecting resistive contributions. The total capacitance is given by  $C_\mathrm{tot} = C_\mathrm{s} + C_\mathrm{p}$, where the sample capacitance $C_\mathrm{s}$ is much smaller than the stray capacitance of the PCB. $C_\mathrm{s}$ is determined by the geometric source and drain capacitance of the QD, as well as their quantum capacitance, which depends on whether or not an unoccupied QD state resides within the bias window. A phase shift of $180^{\circ}$ is observed at the resonance. From the width of the resonance curve, we determine a quality factor of $Q \approx 30$ which is in agreement with the estimate $Q \approx 1/Z_0 \sqrt{L/C_\mathrm{tot}} \approx 29$ (line impedance $Z_0 = 50~\Omega$) and with earlier work using surface mount inductors~\cite{West2019May}.

In order to form a single-electron QD in the BLG sheet, we first define a narrow conductive channel by opening a displacement field induced band gap underneath the split gates. We achieve this by applying a constant back gate voltage of $V_\mathrm{BG} = -3.3 \, \mathrm{V}$ and a split gate voltage of $V_\mathrm{SG}= 2.52 \, \mathrm{V}$ resulting in a band gap of around 55~meV in the regions below the SGs and an overall p-doped channel~\cite{McCann2006}. 
Second, we apply a positive voltage to one of the finger gates to locally tune the band edges of the gapped BLG with respect to the Fermi level. At sufficiently large $V_\mathrm{FG}$, the conduction band edge is pushed below the Fermi level, leading to the formation of an electron QD as illustrated in Fig.~\ref{f1}(c).
The formation of a QD is verified by a measurement of the source-drain current through the device as a function of $V_\mathrm{FG}$. Fig.~\ref{f1}(d) shows Coulomb peaks of the QD in the single to few electron regime. The peaks are grouped in quadruplets in agreement with the spin and valley degeneracy in BLG QDs~\cite{Eich2018Aug,Banszerus2020Jun,Garreis2020Nov}. 

Fig.~\ref{f1}(e) shows the phase shift of the reflected  RF signal at resonant excitation as a function of $V_\mathrm{FG}$. 
The peak positions correspond to the Coulomb peaks measured in DC transport. 
We observe a phase shift of $\Delta\phi \approx 0.4^\circ$ at the first charge transition compared to the Coulomb blocked regime. The phase shift is caused by a shift of the resonance frequency, as the quantum capacitance changes if electrons can tunnel in and out of the QD. Hence, this method is sensitive to charge transitions in contrast to techniques relying on capacitively coupled detectors sensing absolute charge. We can estimate the shift in quantum capacitance according to $\Delta C \simeq  \Delta\phi \cdot C_\mathrm{tot}/2Q \approx 45~$aF~\cite{Chorley2012Jan,Colless2013Jan}.

Next, we perform finite bias spectroscopy measurements in the few electron regime. Fig.~\ref{f2}(a) and (b) compare data sets obtained from DC transport and from dispersive readout. The fourfold shell filling pattern can be identified from the Coulomb diamond measurements. We determine a charging energy of the first electron transition of $E_\mathrm{C} = 6~$meV which corresponds to a diameter of the QD of $\approx 150$~nm according to a simplified plate capacitor model. This is in reasonable agreement with the lithographic dimensions of the device. Outside the regions of Coulomb blockade, a rich spectrum of excited states can be observed. A comparison of the two data sets in Fig.~\ref{f2} shows that dispersive readout is sufficiently sensitive to resolve individual excited state transitions as observed in transport.
The readout technique remains sensitive over the entire gate voltage span, which is in contrast to capacitively coupled QPC and SET charge sensors typically requiring additional compensation voltages to keep them at maximum sensitivity.

In the following, we characterize the signal quality more quantitatively.
Fig.~\ref{f3}(a) shows the signal-to-noise ratio (SNR) of the phase response at the first Coulomb peak as a function of the RF frequency, $f$. 
We define the SNR as the ratio of the phase shift $\Delta\phi$ and the root mean square of the phase noise $\langle \phi \rangle$ in the Coulomb blocked regime. A maximum SNR of $\approx 7$ is observed at resonance ($f = 279$~MHz and RF power of -95~dBm). 
The SNR as function of excitation frequency exhibits a full width at half maximum of approx. 4~MHz.
Fig.~\ref{f3}(b) shows the phase shift as a function of $V_\mathrm{FG}$ for different RF power at resonant excitation. At an RF power at the sample below $-100$~dBm, the low signal amplitude causes a significantly increased noise floor making the Coulomb peaks hardly detectable. At high RF power, dissipation in the device results in a thermal broadening of the Coulomb peaks which can be observed above $-84$~dBm and leads to almost fully vanishing peaks at $-78$~dBm. For a quantitative analysis, the SNR obtained from the first Coulomb peak as a function of excitation power is plotted in Fig.~\ref{f3}(c). The SNR is maximised at around $-85$~dBm and is reduced by the aforementioned effects.

In order to study the temporal resolution of the implemented dispersive charge sensing technique, we plot the SNR as function of the bandwidth of the demodulation circuit (see Fig.~\ref{f4}). While a SNR of $\approx17$ is observed for a frequency of $f=3$~Hz, we determine the maximum bandwidth of the sensor to be on the order of 1~kHZ, where the SNR reaches unity. Following the definition of the sensitivity in Ref.~\cite{Zheng2019Aug}, we obtain $e\sqrt{t_\mathrm{min}}\approx3.1\times10^{-2}~e/\sqrt{\mathrm{Hz}}$, where $t_\mathrm{min}$ is the integration time where the SNR reaches unity. This sensitivity is comparable to similar results for dispersive charge sensing on III-V semiconductor quantum dots~\cite{Colless2013Jan,Petersson2010Aug}.

In conclusion, we demonstrated dispersive charge sensing on a single-electron BLG QD. We use a RF reflectometry readout scheme to detect the shift in quantum capacitance caused by the QD being on and off resonance. We are able to resolve the excited state spectrum of a QD in the few electron regime. 
Omitting capacitively coupled mesoscopic charge detectors brings significant advantages as it simplifies constraints in device design, reduces the risk of leakage currents and prevents back action caused by shot noise~\cite{Gustavsson2007Nov} generated in the detector. Furthermore, the dispersive readout technique remains sensitive over a wide gate voltage range without the need for compensation voltages to retain sensitivity.
However, our measurements show limitations in the measurement bandwidth and point towards further improvements of the RF circuitry: The $Q$-factor of the resonant circuit can be increased by the use of superconducting on-chip inductors minimizing resistive losses~\cite{Colless2013Jan,Hornibrook2014Mar}. Parasitic capacitances can be reduced by design optimization of the PCB, further enhancing the sensitivity on the changes in quantum capacitance, ultimately allowing for single shot detection~\cite{West2019May}.
Nevertheless, our measurements show the high potential of dispersive charge sensing for the readout of charge and spin states in bilayer graphene QDs.\\

\textbf{Acknowledgements} The authors thank S.~Trellenkamp and  F.~Lentz for their support in device fabrication.
This project has received funding from the European Union's Horizon 2020 research and innovation programme under grant agreement No. 881603 (Graphene Flagship) and from the European Research Council (ERC) under grant agreement No. 820254, the Deutsche Forschungsgemeinschaft (DFG, German Research Foundation) under Germany's Excellence Strategy - Cluster of Excellence Matter and Light for Quantum Computing (ML4Q) EXC 2004/1 - 390534769, through DFG (STA 1146/11-1), and by the Helmholtz Nano Facility~\cite{Albrecht2017May}. Growth of hexagonal boron nitride crystals was supported by the Elemental Strategy Initiative conducted by the MEXT, Japan, Grant Number JPMXP0112101001,  JSPS KAKENHI Grant Numbers JP20H00354 and the CREST(JPMJCR15F3), JST.

\bibliography{literature}
\clearpage

\end{document}